\documentclass[natbib]{svjour3}            
\smartqed  
\usepackage{graphicx}
\usepackage{color}
\usepackage{epsfig,amsmath,natbib}
\newcommand{\NL}{nonlinear}
\newcommand{\DSA}{diffusive shock acceleration}
\newcommand{\MFA}{magnetic field amplification}
\newcommand{\BellAmp}{N_B} 

\newcommand{\vkvec}{\mathbf{v}(\mathbf{k})}
\newcommand{\bkvec}{\mathbf{b}(\mathbf{k})}
\newcommand{\hydro}{hydrodyamical}

\newcommand{\Tmax}{\theta_\mathrm{max}}

\newcommand{\vph}{v_{\rm ph}}

\newcommand{\JzCR}{J_0^\mathrm{cr}}

\newcommand{\nonres}{nonresonant}
\newcommand{\GamMax}{\gamma_\mathrm{max}}

\newcommand{\Ppar}{P_{\parallel}}
\newcommand{\Pperp}{P_{\perp}}

\newcommand{\rgz}{r_{g0}}
\newcommand{\JCR}{\mathbf{J}^\mathrm{cr}}

\newcommand{\nCR}{n_\mathrm{cr}}

\newcommand{\syn}{synchrotron}
\newcommand{\synch}{synchrotron}

\newcommand{\rel}{relativistic}
\def\lsim{\;\raise0.3ex\hbox{$<$\kern-0.75em\raise-1.1ex\hbox{$\sim$}}\;}
\def\gsim{\;\raise0.3ex\hbox{$>$\kern-0.75em\raise-1.1ex\hbox{$\sim$}}\;}

\newcommand{\la}{\raise0.3ex\hbox{$<$}\kern-0.75em{\lower0.65ex\hbox{$\sim$}}}
\newcommand{\ga}{\raise0.3ex\hbox{$>$}\kern-0.75em{\lower0.65ex\hbox{$\sim$}}}

\def\lsim{\;\raise0.3ex\hbox{$<$\kern-0.75em\raise-1.1ex\hbox{$\sim$}}\;}
\def\gsim{\;\raise0.3ex\hbox{$>$\kern-0.75em\raise-1.1ex\hbox{$\sim$}}\;}
\def\alf{Alfv\'en }

\def\etal{{ et al. }}

\journalname{Space Science Reviews}
\begin{document}

\title{Magnetic fields in cosmic particle acceleration sources}

\titlerunning{Magnetic fields in CR sources} 

\author{Andrei M. Bykov \and Donald C. Ellison \and Matthieu Renaud }

\authorrunning{Bykov et al.} 

\institute{Andrei Bykov \at A.F.Ioffe Institute for Physics and
Technology, 194021 St. Petersburg, Russia,
\\ also St. Petersburg State Politechnical University \email{byk@astro.ioffe.ru} \\ \and Donald C. Ellison \at Physics
Department, North Carolina State University, Box 8202, Raleigh, NC
27695, USA \email{don\_ellison@ncsu.edu}\\ \\ \and Matthieu Renaud
\at Laboratoire de Physique Theorique et Astroparticules (LPTA)
Universite Montpellier II, France \email{mrenaud@lpta.in2p3.fr} }

\date{Received: date / Accepted: date}

\maketitle

\begin{abstract}
We review here some magnetic phenomena in astrophysical particle
accelerators associated with collisionless shocks in supernova
remnants, radio galaxies and clusters of galaxies. A specific
feature is that the accelerated particles can play an
important role in magnetic field evolution in the objects. In
particular, we discuss a number of cosmic-ray (CR) driven, magnetic
field amplification processes that are likely to operate when
diffusive shock acceleration (DSA) becomes efficient and nonlinear.
The turbulent magnetic fields produced by these processes determine
the maximum energies of accelerated particles and result in specific
features in the observed photon radiation of the sources.  Equally
important, magnetic field amplification by the CR currents and
pressure anisotropies may affect the shocked gas temperatures and
compression, both in the shock precursor and in the downstream flow,
if the shock is an efficient CR accelerator. Strong fluctuations of
the magnetic field on scales above the radiation formation length in
the shock vicinity result in intermittent structures observable in
synchrotron emission images. The finite size twinkling, intermittent
structures -- dots, clumps, and filaments -- are most apparent in
the cut-off region  of the synchrotron spectrum. Even though these
X-ray \syn\ structures result from turbulent magnetic fields, they
could still be highly polarized providing an important  diagnostic
of the spectrum of the turbulence. We discuss both the thermal and
non-thermal observational consequences of \MFA\ in supernova
remnants and radio-galaxies.
Resonant and non-resonant CR
streaming instabilities in the shock precursor can generate mesoscale
magnetic fields with scale-sizes comparable to supernova remnants and
even superbubbles. This opens the possibility that magnetic fields in
the earliest galaxies were produced by the first generation Population III supernova remnants
and by clustered  supernovae in star forming regions.

\end{abstract}

\section{Introduction}
Particle acceleration takes place in many active astrophysical
objects of very different nature and scales. Magnetic fields play
the central role in charged particle acceleration either as a
intermediary between the plasma flows and energetic particles, as
is the case for Fermi acceleration (e.g., in collisionless shocks), or
as a source of free energy to be converted into energetic
particles (i.e., magnetic field reconnection processes).

The existence of the highly amplified magnetic fields in the range of
0.1-1 mG in the shells of young supernova remnants (SNRs) was
established assuming equipartition between relativistic particles and
magnetic fields in the synchrotron radio emitting shells \citep[see
e.g.,][and the references therein]{gs64}. The topology of the magnetic
field in SNRs inferred from the observations of synchrotron emission
differ strongly between young and old SNRs.  Radio polarization studies
reveal super-adiabatic magnetic field amplification and a net radial
orientation of the magnetic fields in young SNRs, while it is often just
shocked interstellar field, mainly tangential, in old SNRs
\citep[e.g.,][]{milne90}. Milne has also pointed out that as the
resolution increases, the polarization structure becomes more complex.

A pixel-by-pixel map of Faraday rotation has been
produced  applying rotation measure (RM) synthesis to the data
observed with the \emph{Australia Telescope Compact Array}, for the
entire supernova remnant  G296.5+10.0 by \citet{harveysmithea10}. A
highly ordered rotation measure structure, with an anti-symmetric
rotation measure pattern, was observed.
The authors propose that the observed rotation measures are the imprint
of an azimuthal magnetic field in the stellar wind of the progenitor
star.  A swept-up magnetized wind from a red supergiant can produce an
azimuthal pattern of the magnetic field at large distances from the star
and can naturally produce the observed anti-symmetric RM
pattern. Supernova expansion into such a wind could account for the
apparent bilateral structure of the SNR's radio and X-ray
morphologies. In the case of SN1006, a comparison between observed and
synthesized radio maps, making different assumptions about the
dependence of electron injection efficiency on the shock obliquity,
allowed \citet{petrukea09} to constrain possible nonthermal electron
injection models.

A number of possible amplification mechanisms were considered.
Rayleigh�Taylor instabilities in the shell have been considered for
many years as a potential source of the turbulent magnetic fields.
The Rayleigh-Taylor instability at the interface of the ejecta
and the shocked ambient medium was proposed to explain these
observations.
\citet{jn96}  performed multi-dimensional MHD simulations of the
instability in the shell  and its effect on the local magnetic
field. They found that the evolution of the instability is very
sensitive to the deceleration of the ejecta and the evolutionary
stage of the remnant. The Rayleigh-Taylor and Kelvin-Helmholtz
instabilities amplify ambient magnetic fields in the simulations
locally by as much as a factor of 60 around dense fingers due to
stretching, winding, and compression. Globally, the amount of
magnetic-field amplification was nevertheless low and the magnetic
energy density reaches only about 0.3\% of the turbulent energy
density at the end of simulation. Strong magnetic field lines draped
around the fingers produce the radial B-vector polarization.
Magnetic Rayleigh-Taylor instabilities in three dimensions were
simulated by \citet[][]{sg07}, with a focus on the nonlinear structure
and evolution that results from different initial field
configurations. They found that strong magnetic fields do not
suppress the instability but reduce mixing between the heavy and light
fluids and cause the rate of growth of bubbles and fingers to
increase in comparison to hydrodynamics.
\citet{schureea08} suggested that higher compression ratios and
additional turbulence due to the dominant presence of CRs can be
important in attempts to reproduce the observed magnetic field
morphology with the Rayleigh-Taylor instability simulations. Both
numerical simulations and laboratory laser experiments are ongoing to
study the Rayleigh�Taylor instability problem. The possible effects
of magnetic fields on laser experiments of Rayleigh�Taylor
instabilities in supernova-like setups was discussed recently by
\citet{fryxellea10}. Comparing numerical simulations, as well as
experimentally using the Omega Laser at the University of Rochester, the
authors discussed a possibility that the morphology of the instability
is significantly altered by the generation of very strong magnetic
fields during the laser experiments.

On the other hand, \citet{chevalier77} shown that radio observations of
Tycho's supernova remnant indicated the presence of a collisionless
shock wave undergoing turbulent magnetic field amplification by a factor
of about 20. He noted that the amplification resembled some phenomena in
heliospheric collisionless shocks. In fact, direct support for the
self-generation of MHD turbulence has been seen in heliospheric shocks,
e.g., \citet[][]{bamertea04,bamertea08}. In the upstream region of a
shock associated with the Bastille Day coronal mass ejection, the
excitation of hydromagnetic waves with power spectral density levels
well above the levels in the ambient solar wind were observed by
\citet{bamertea04} with the magnetometer on board the \emph{Advanced
Composition Explorer}.  \citet{abr94} set upper limits on the scattering
mean free paths of radio emitting electrons in front of
supernova remnant shock fronts using high-resolution radio observations of four
Galactic supernova remnants. They found that, for the sharpest
synchrotron radio rims, the mean free path is typically less than one
percent of that derived for cosmic rays of similar rigidity in the
interstellar medium,  implying that enhanced hydromagnetic wave intensity
is generated by diffusive shock acceleration.

Recent observations of X-ray \synch\ radiation from several young SNRs
  revealed thin X-ray filaments in the vicinity of blast waves. This
  provides indirect, but rather strong evidence for strong
  super-adiabatic magnetic field amplification (MFA) in the blast wave
  associated with CR production \citep[see
  e.g.,][]{vl03,bambaea05,bambaea06,uchiyamaea07,pf08,reynolds08,vink08}.
The synchrotron filamentary structures could be due to a narrow
spatial extension of the highest energy electron population in the
shock acceleration region limited by rapid synchrotron energy
losses. In this case, strong magnetic field amplification in the
shock vicinity is required to match the observed thickness of the
non-thermal filaments \citep[e.g.,][]{vl03,bambaea05}, thus
supporting the case for highly efficient DSA. An alternative
interpretation, that the observed narrow filaments are limited by
magnetic field damping and not by the energy losses of the radiating
electrons, has also been proposed
\citep[see][for example]{pyl05,buttea08}.
\citet{CassamEtal2007,Cassam2008} have studied this effect in detail
for Tycho's SNR and
  SN1006.

Diffusive shock acceleration is a very promising mechanism for
producing superthermal and \rel\ particles in a wide range of
astrophysical objects ranging from the Earth bow shock \citep[e.g.,][]{EMP90} to shocks in
galaxy clusters \citep{be87,je91,md01,bdd08}. This mechanism is
believed to be efficient \citep[e.g.,][]{helderea09} and capable of
producing CRs to energies well above $10^{15}$\,eV in
young SNRs \citep[e.g.,][]{pzs10}, and even higher in active
radio-galaxies such as Cen A \citep[see, for
example,][]{crostonea09}.

Fast and efficient DSA demands that particles are multiply scattered by magnetic fluctuations in the shock vicinity. The amplitude of the
required MHD turbulence is substantially higher than the ambient MHD
turbulence suggesting that the turbulence is generated by the shock.
%
In order for DSA to accelerate
particles to high energies, the energetic particles must be able to
interact with magnetic turbulence over a broad wavelength range. The
weakly anisotropic distribution of accelerated particles,  is believed capable of producing this turbulence
in a symbiotic relationship where the magnetic turbulence required
to accelerate the CRs is created by the accelerated CRs themselves.
In efficient DSA, this wave-particle interaction can be strongly
nonlinear where CRs modify the plasma flow and the specific
mechanisms of magnetic field amplification.

\section{Magnetic field amplification by CR-driven
instabilities}\label{mfa}
The non-equilibrium distributions of
 accelerated particles in efficient
DSA are believed capable of producing magnetic turbulence and
%
magnetic field amplification.
For the forward shock in a young, isolated SNR, the
accelerated CR pressure can be a sizeable fraction of the shock ram
pressure and far exceed the background plasma pressure. The anisotropic
distribution of CRs in the shock precursor results in ponderomotive
forces on the background plasma that can cause the amplification of
magnetic fluctuations in a certain wavevector range and polarization. We
briefly discuss in section \ref{anis} the kinetic background for the
description of this CR anisotropy and then outline a number of
instabilities related to the anisotropic CR distributions.

\subsection{Anisotropic CR distribution in DSA}\label{anis}
First-order Fermi acceleration (also called diffusive shock
acceleration) of CRs by non-relativistic magnetized flows,
is characterized by a nearly
isotropic CR momentum distribution in a broad energy range. Two notable
exceptions to this are injected, barely superthermal particles, and
the highest energies CRs.
The CR distribution is formed by the multiple scattering of charged
particles across the shock by magnetic fluctuations over a wide dynamic range. Cosmic-ray
particle dynamics in DSA is governed by electro-magnetic fields both
regular and stochastic. The distribution function of energetic
particles, $f(\mathbf{r},\mathbf{p},t) = \langle
F(\mathbf{r},\mathbf{p},t) \rangle_c$, averaged over electro-magnetic
fluctuations with scales below some value $l_c$, appropriate for a
particular problem, satisfies the kinetic equation
\begin{equation}\label{kin_eq}
\frac{\partial f}{\partial t}+\mathbf{v}\cdot\frac{\partial
f}{\partial\mathbf{r}}+e\mathbf{E}\cdot\frac{\partial
f}{\partial\mathbf{p}}-\frac{ec}{\mathcal{E}}\mathbf{B}
\cdot\widehat{\mathbf{\mathcal{O}}}f=\widehat{I}_c\,[f],
\end{equation}
where $\widehat{\mathbf{\mathcal{O}}}$ is the momentum rotation
operator, defined by
\begin{equation}\label{a5}
\widehat{\mathbf{\mathcal{O}}}=\mathbf{p}\times\frac{\partial
}{\partial\mathbf{p}} \ ,
\end{equation}
$\mathcal{E}$ is the particle energy, and $\mathbf{E} = \langle
\mathbf{E}_0\rangle_c$ and $\mathbf{B}= \langle
\mathbf{B}_0\rangle_c$ are the electric and magnetic fields averaged
over electro-magnetic fluctuations with scales below $l_c$.
The fluctuating parts of the
electromagnetic fields are $\mathbf{b}= \mathbf{B}_0 -
\langle \mathbf{B}_0\rangle_c$ and  $\mathbf{e}= \mathbf{E}_0 - \langle
\mathbf{E}_0\rangle_c$.
The fields are self-consistent since the sources in Maxwell's equations
include the CR currents determined by $F(\mathbf{r},\mathbf{p},t)$. We
should stress here that it is not trivial to get a closed system of
equations for the averaged distribution function
$f(\mathbf{r},\mathbf{p},t) = \langle F(\mathbf{r},\mathbf{p},t)
\rangle_c$ since the higher order moments of $F$ may contribute through
$\langle \mathbf{b}(\mathbf{r}, t) F(\mathbf{r},\mathbf{p},t) \rangle_c$
and $\langle \mathbf{e}(\mathbf{r}, t) F(\mathbf{r},\mathbf{p},t)
\rangle_c$. In fact, even the existence of a well defined averaging
procedure, and the averages themselves,  is assumed rather than proved.

The collision operator
\begin{equation}\label{coll_oper}
\widehat{I}_c\,[f] = - Ze\langle \mathbf{e}\cdot\frac{\partial
F}{\partial\mathbf{p}}-\frac{c}{\mathcal{E}}\mathbf{b}
\cdot\widehat{\mathbf{\mathcal{O}}}F\rangle_c,
\end{equation}
is, in general, a nonlinear operator if the effect of energetic
particles on the electromagnetic fields  is non-negligible.

Different approaches are used to get a closed analytic expression
for $\widehat{I}_c\,[f]$ including Fokker-Planck type equations
and quasi-linear theory \citep[see
e.g.,][]{toptygin85,be87,berezinskiiea90,schlikeiser02}.
The simplest case for the collision operator can be deduced from
Eq.~(\ref{coll_oper}) for particles scattered by magnetic fluctuations
with scales, $l_c$, smaller than the particle gyroradius, $r_{g} =
cp/eB$, in the total magnetic field B (random plus coherent).
In the high-energy limit, $r_{g}>>l_c$, the particle momentum change
over the field correlation length $l_c$ is small ($\propto l_c/r_{g}$) and
the scattering rate in Eq.~(\ref{coll_oper}) reduces to $v/2\Lambda(p)
\,\,\widehat{\mathbf{\mathcal{O}}}^2$ with $\Lambda(p) =
\rm{g}\,R_{\rm{st}}^2(p)/l_c$. Here, $R_{\rm{st}}(p)$ is the particle
gyroradius in the stochastic magnetic field, and $g$ is a numerical factor depending on the short-scale fluctuation spectral index. The
strong momentum dependence, $\Lambda \propto p^2$, makes this scenario
unfavorable for confining and accelerating high-energy particles in DSA.

It is important to note that $\widehat{\mathbf{\mathcal{O}}}^2$ can be
reduced to the angular part of the Laplace operator with well-known
spherical harmonics as the eigenfunctions. The high-energy asymptotic
range is valid even in the case of strong (but quasi-static) magnetic
fluctuations of amplitudes larger than the mean magnetic field.
However, for particles of intermediate and low energies, with $r_{g}
\lsim l_c$ and scattered by strong magnetic fluctuations, the collision
operator is no longer simple and only some phenomenological scheme like
the \emph{relaxation time approximation} can be used. These schemes do
not have a microscopic justification.

The momentum distribution of accelerated particles,
$f(\mathbf{r},\mathbf{p},t)$, of a weakly anisotropic distribution
can be presented as
\begin{equation}\label{distrFuncCr}
f(\mathbf{p})= \frac{\nCR}{4\pi}N(p)\left[1+\delta_{\rm
a}(\mathbf{p})\right],
\end{equation}
where $\nCR$ is the CR number density, $N(p)$  is the isotropic and
$\delta_{\rm a}(\mathbf{p})$ is the anisotropic part of the
distribution function defined to satisfy
$$
\int{\delta_{\rm a}(\mathbf{p})}\,d\Omega_{\mathbf{p}} = 0,~~
\int{f(\mathbf{p})}\,p^2dp\,d\Omega_{\mathbf{p}} = \nCR
\ .
$$ Due to CR particle scattering by magnetic fluctuations of different
scales, the distribution function is nearly isotropic on scales larger
than the CR mean free path, $\Lambda(p)$, and, therefore, its anisotropic
part is small, i.e., $\delta_{\rm a}(\mathbf{p}) \ll $ 1 on scales $k
\Lambda(p) <1$. The actual angular dependence of $\delta_{\rm
a}(\mathbf{p})$, and the momentum dependence of $\Lambda(p)$, are
determined by the exact form of the collision operator, the structure of
regular fields, and the boundary conditions. It can be presented in the
spherical harmonic expansion
%
%
\begin{equation}\label{expans}
\delta_{\rm a}(\mathbf{p}) =
\sum_{l,m}\Delta_{l,m}(p)\,Y_{l,m}(\vartheta,\varphi),~~~~~ |m|\leq
l,
\end{equation}
where $\vartheta$ and $\varphi$ are the pitch and azimuthal angles
of the particle momentum, correspondingly. Since the isotropic part
of the distribution function is separated from the anisotropic part
in Eq.~(\ref{distrFuncCr}), $\Delta_{0,0} = 0$. The current-type
anisotropy that is determined by $\Delta_{1,m}<1$, is usually
considered to be the dominant term. That results in the standard
diffusion approximation and the transport equation for the isotropic
part $N(\mathbf{r}, p, t)$. In the case of a high-energy particle
scattering off of small-scale fluctuations, $r_{g} \gg l_c$, the
spherical functions are the eigenfunctions of the collision operator
since $\widehat{\mathbf{\mathcal{O}}}^2\,Y_{l,m}=-l(l+1)Y_{l,m}$. In
that case, indeed, $\Delta_{2,m} \sim \Delta_{1,m}^2$. However, the
high multipole components of the distribution can be substantial in
the case of anisotropic scattering, or in plasma flows with a
complex structure of quasi-regular magnetic fields (e.g., in the
presence of regular or stochastic magnetic traps).
In the
latter case one can expect $\Delta_{2,m} \sim \Delta_{1,m}$ and the
transport equation for the isotropic part of the distribution
function $N(\mathbf{r}, p, t)$ can differ from the standard
advection-diffusion equation that is used in the analytic DSA
models.

Standard analytic DSA models use a simplified description of particle
diffusion. Generally, the
diffusion approximation is made where it is assumed that
a small anisotropy of the
momentum distribution can be completely described by the CR
current and higher moments of the momentum
angular distribution can be neglected
\citep[see e.g.,][]{be87,je91,md01}.
Then, further assuming that magnetic fields are frozen-in the plasma, the kinetic
Eq.~(\ref{kin_eq}) can be reduced to a transport equation for the
isotropic part of the distribution function $N(\mathbf{r}, p,t)$.
This approach can account for particle advection, diffusion and
acceleration by the first-order Fermi mechanism in
non-relativistic shocks. The diffusion coefficient in these analytical
models is parameterized as $D(p) \propto r_g^{\rm{a}}$, where the
numerical index is generally in the range
$0\leq\rm{a}\leq2$. The widely used phenomenological
\emph{Bohm diffusion model} assumes $\rm{a}=1$, i.e., the mean free
path  $\Lambda(p) = \eta\, r_g(p)$, where the factor $\eta \geq 1$.

The CR current and anisotropy in the rest frame of a non-relativistic
shock is $|\Delta_{1,m}| \sim v_{s}/c << 1$, where $v_{s}$ is the shock
velocity.  The CR anisotropy can also be parameterized by the CR drift
velocity $v_{\rm drift}$ relative to the background plasma in the rest
frame of the upstream plasma.  Various parameterizations of the CR
current in DSA were used in recent studies of magnetic field
amplification mechanisms by \citep[e.g.,][]{bell04,plm06,ab09,lm09}
based on the CR transport equation. Monte Carlo simulations of DSA
provide an alternative approach to the kinetic theory.  In Monte Carlo
techniques, particle scattering rates are parameterized as a function of
energy and then the CR current (and other CR anisotropy moments) are
derived in nonlinear shock models accounting for the CR backreaction on
the plasma flow \citep[e.g.,][]{ebj96,veb06,vbe08}.

\subsection{Resonant CR-streaming instability}\label{stream}
Resonant interactions have long been known to amplify magnetic
fluctuations on the scale of the CR gyroradius
\citep[e.g.,][]{wentzel74,zweibel03,kulsrud05}.  The simplest case
of interest for DSA is \alf wave amplification by streaming CRs. If
the streaming speed of the cosmic rays through the ambient medium
exceeds the \alf speed, then the amplitude of resonant \alf waves
grows exponentially with time.
The waves are emitted by the  CR
particles gyrating with gyrofrequency $\Omega$ as coherent cyclotron
radiation.  The resonance condition is $k \,v\cos(\vartheta) =
\Omega$. This resonance condition is typical for DSA. The growth
rate is positive if, in the wave frame, the particles stream in the
opposite direction to the motion of the background fluid.
Equivalently, the waves grow if, in the fluid frame, the mean
streaming velocity of the particles exceeds the wave velocity, i.e.,
$v_{\rm drift}> \vph$. The growth rate estimated by \citet{be87} is
\begin{equation}\label{res}
\gamma^{\rm res} \approx \Omega\, \left[\frac{v_{\rm drift}}{\vph} -
1\right] \frac{n_{\rm cr}}{n_{\rm a}} \ ,
\end{equation}
where $n_{\rm cr}$ is the total number density
of resonant cosmic rays and
$n_{\rm a}$ is the background ion density.

\subsection{Non-resonant streaming instabilities: short-scale}\label{bell}
In addition to resonant amplification, the CR current can also efficiently
amplify non-resonant magnetic fluctuations with scales smaller than
the CR gyroradius \citep{bell04}. The following two points are important
to produce strong, short-scale turbulence.

(i) The cosmic-ray current has only a weak response to the fluctuations
 with scales shorter than the gyroradii, $r_{g0}$, of CR particles
 dominating the CR current (short-scale fluctuations), i.e., $kr_{g0} >>
 1$ (see discussion in section \ref{anis}). In this case, the electric
 current of accelerated particles $\JCR$ is an external current in the
 momentum equation of the background plasma. The current is governed by
 sources of energetic particles and by local electromagnetic fields. The
 current $\JCR$ initiates a compensating return current in the
 background plasma.

(ii) The ponderomotive force of the CR current, $\JCR \times
\mathbf{B}/c$, is large enough to dominate the magnetic field tension
in the momentum equation of the background plasma. Then the $\JCR
\times \mathbf{B}/c$ force from the CR current drives the Bell
short-scale instability at scales below the CR gyroradius. This
condition is satisfied for scales $k < k_1$ where
\begin{equation}\label{k1}
k_1 = \frac{4\pi}{c} \frac{\JzCR}{B_0} \ .
\end{equation}
Note that the condition for mode growth, $\rgz k > 1$, together
with Eq.~(\ref{k1}), implies that the CR current that determines
$k_1$ must be large enough to drive the non-resonant instability.
The anisotropy of the relativistic particle distribution,
$|\Delta_{1,0}|$, should exceed the ratio of the mean magnetic field
energy density to the energetic particle energy density
$E_\mathrm{cr}$. That is
\begin{equation}\label{aniz}
|\Delta_{1,0}| > \frac{B^2}{4 \pi E_\mathrm{cr}},
\end{equation}
where the CR current is given by $\JzCR \approx |\Delta_{1,0}| e
n_{cr} c$.

\citet{bell04, bell05} discovered that the system is unstable
against linear perturbations that are
$\propto \exp(\gamma t+i\mathbf{k}\mathbf{r})$
and the return current drives nearly
purely growing modes. Here, $\gamma$ is the linear growth rate.

In a cold plasma with sound speed $a_{0}$, well below the \alf
velocity $v_{a}$, the linear growth rate obtained by \citet{bell05}
depends only on the wavevector projection $k_{z}$ on the local mean
magnetic field, i.e.,
\begin{equation}\label{DispU}
\gamma\approx \GamMax k_z/k ,
\end{equation}
where
\begin{equation}\label{DispU1}
\GamMax = v_{a}\sqrt{k_{1}|k|-k^{2}}
\ .
\end{equation}

According to the linear analysis of \citet{bell05}, the wavenumber
of a growing mode must satisfy the condition $\rgz^{-1}< k <
k_{1}$. Therefore, the instability growth rate is higher than the
\alf frequency $v_{a}k$. The amplitudes of the growing mode in the
linear approximation can be expressed as
\begin{equation}\label{b2v2}
|\vkvec|^{2}\approx
\frac{1}{4\pi\rho}\frac{k_{1}}{|k_{z}|}|\bkvec|^{2},
\end{equation}
provided that the kinetic energy density in the growing mode
dominates over the magnetic energy density because $k_{1}> k_{z}$.
This is in contrast to \alf modes where the energy densities are
equal.

The cold plasma approximation used in the analysis of Bell turbulence is
valid if
\begin{equation}\label{thermEf}
\left(\frac{v_{a}}{v_{Ti}}\right)^{2} > k_{1}r_{g0}\frac{v_{a}}{c},
\end{equation}
where $v_{Ti}$ is the thermal ion velocity.
The approximation is typically good for galactic SNRs in a warm
interstellar medium where the plasma parameter $\beta = a_0^2/v_a^2
\sim 1.0$. In the case of DSA by large-scale shocks in superbubbles
\citep[see e.g.,][]{bt93,bt01}, or in hot intercluster plasma,
the thermal corrections to the wave dispersion relations can be
essential. A thorough discussion of the effects of a hot plasma on
the short-scale modes was given by \citet{ze10}.

\emph{Particle-in-cell} simulations by \citet{rs09,rs10} showed that
the back-reaction of the amplified field on CRs would limit the
amplification factor of the Bell instability to less than a factor
of about 10 in the upstream flows of galactic SNRs.
The authors studied
the possibility of further amplification driven near shocks by
magnetized CRs, whose Larmor radii are smaller than the length scale
of the field that was previously amplified by the Bell instability in the upstream flow.
They found that additional amplification can occur due to the CR
current perpendicular to the field. The maximum amplification of the
instability is determined by the disruption of the CR current, which
happens when CR Larmor radii in the amplified field become
comparable to the length scale of the instability. Amplification
factors up to $\sim 45$ were estimated in that case.

An important feature of the instability driven by the perpendicular
current established by \citet{rs10} is the characteristic dependence
of the amplified field on the shock velocity, $B^2 \propto v_s^2$,
which contrasts with the Bell instability acting alone where $B^2
\propto v_s^3$. Different scalings can, in principle, be constrained
in statistical studies of radio SNRs, as done by \citet{bp10},
but the results are sensitive to other parameters such as ambient
density.
It is also possible to constrain these scalings by comparing radio
  and X-ray \syn\ lightcurves of individual young supernova shells. In
  an amplified magnetic field, the \syn\ lightcurve will differ from the
  radio if strong magnetic field damping behind the shock occurs.

\subsection{Non-resonant streaming instabilities: long-wavelength}\label{longwave}
The nonresonant, short-scale instability introduced
by \citet{bell04}, and
discussed in section \ref{bell}, is fast and can strongly amplify
short-scale magnetic field fluctuations. These strong, short-scale
fluctuations are important for the dynamics of CR modified shocks and
their emission properties, however, to study the maximum energies
of CR particles in DSA one needs to study the amplification
mechanisms for long-wavelength fluctuations of scales  $\rgz k <
1$.
Specifically, what are the consequences of long-wavelength fluctuations
propagating in highly turbulent plasma with much shorter scale
fluctuations? Also, contrary to the short-scale case, the response of
the magnetized CR current on magnetic field fluctuations is not small in
the limit $\rgz k < 1$.

Using a multi-scale, quasi-linear analysis, \citet{boe10} showed that
the presence of turbulence with scales shorter than the CR
gyroradius
  enhances the growth of modes with scales longer than the gyroradius,
  at least for particular polarizations.  They used a mean-field approach
  to average the equation of motion and the induction equation over the
  ensemble of magnetic field oscillations accounting for the anisotropy
  of relativistic particles on the background plasma.
\citet{boe10} derived the response of the magnetized CR current on magnetic
field fluctuations and showed that, in the presence of short-scale
Bell-type turbulence, long wavelength modes are amplified.
In general, the growth rates depend on the mode propagation angle. For modes
propagating parallel to the initial magnetic field, the growth rate
is
\begin{equation}\label{incr2}
\gamma^{lw}(k) \approx
\sqrt{\frac{\pi\BellAmp}{2}}\sqrt{\frac{kk_{0}}{\eta}}v_{a} \ ,
\end{equation}
and these modes have the fastest growth rates for the Bohm diffusion
regime with $\eta \sim 1$.

The propagation angle $\Tmax$ of the mode of maximum growth for
$\eta>1$ is
\begin{equation}\label{thetaMax}
\cos{\Tmax} = 1 / \eta ,
\end{equation}
and the maximum growth rate at $\eta\gg 1$ is determined by
\begin{equation}\label{incr8}
\gamma(k)\approx\sqrt{\frac{\pi\BellAmp}{4}}\sqrt{kk_{0}}v_{a}.
\end{equation}

\begin{figure*}
\includegraphics[width=1\textwidth]{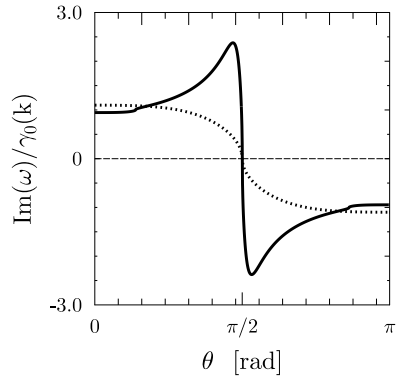}
\caption{The angular dependence of the growth rates of the
long-wavelength unstable modes in the \hydro\ regime where $kr_{g0}
< \eta^{-1}$, for $\eta = 10$. The angle $\theta$ is between
the wavevector and the CR current in the upstream rest frame of a
parallel shock. The direction of the magnetic field determines the
polarization (helicity) of the modes. The model parameters are
$k_{1}r_{g0} = 100$, the particle distribution index $\alpha = 4.0$,
and the plasma parameter $\beta = a_0^2/v_a^2 = 1.0$.  The solid and
dot-dashed curves show the two unstable modes. The normalizing
parameter $\gamma_0$ is determined by Eq.~(\ref{incr2}). See Bykov
et al. (2011) for a comparison with the Bohm diffusion case $\eta =
1$ }\label{angular}\end{figure*}

\begin{figure*}
\includegraphics[width=1\textwidth]{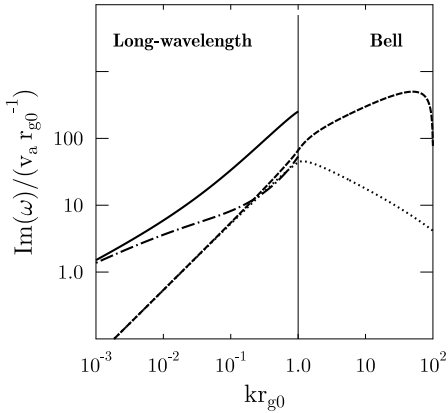}
\caption{The figure, from  Bykov et al. (2010), shows growth rates
of the parallel propagating modes as a function of the wavenumber to
illustrate the effect of short-scale
    turbulence on the long-wavelength instability.
    The model parameters
    are $k_{0}r_{g0} = 100$ and $\alpha = 4.0$. The solid and dot-dashed
    curves are simulated for two modes in the model with short-scale
    turbulence of $\xi =5$ and $\eta = 10$ to demonstrate the behavior
    of the modes in the intermediate regime. For comparison, the dashed
    and dotted curves are calculated for the model without the
    short-scale turbulence, i.e., with $\BellAmp =0$ and $\eta
    \rightarrow \infty$ \citep[c.f.,][]{bell04}.}\label{growth1}
\end{figure*}

The above results were obtained assuming $v_{\rm{ph}}/v_s \ll 1$. In the
case under consideration $\gamma(k)\sim\omega(k)$, and therefore from
Eq.~(\ref{incr8}) one may get the validity condition in the form
$\displaystyle\sqrt{\pi\BellAmp/4}\sqrt{k_0/k}\,M_{\rm{a}}^{-1}
\ll 1$,
%
%
where $M_{\rm{a}}$ is the \alf Mach number of the shock. The angular
dependence of the growth rate in the hydrodynamical regime depends
on the dimensionless collision strength $\eta$, as is shown in
Fig.~\ref{angular} of \citet{boe10}. The growth rates of  resonant,
non-resonant Bell's,  and non-resonant long wavelength instabilities
are illustrated in Fig.~\ref{growth1} taken from \citet{boe10} where
the polarization, helicity, and angular dependence of the growth
rates are calculated for obliquely propagating modes for wavelengths
both below and above the CR mean free path.  The long-wavelength
growth rates estimated for typical supernova remnant parameters are
sufficiently fast to suggest a fundamental increase in the maximum
CR energy a given shock can produce. It should be noted that the
phase velocity of the long-wavelength modes is a growing function of
the wavelength and it is larger than the \alf speed. That may
steepen the particle spectra at the highest energy end by reducing
the effective shock compression at high energies. Also, the high
phase velocity of the modes could increase the role of
Fermi-II acceleration.

An important point is that the 
short-scale turbulence can influence the large-scale dynamics
through the ponderomotive forces imposed on the plasma by the
turbulence and the CR current. To derive the mean ponderomotive
force one must average the momentum equation over the ensemble of
short-scale fluctuations.
When this is done, it is seen that the ponderomotive force due to
the CR current response may result in a long-wavelength instability
in a way somewhat similar to Bell's instability.


\subsection{Non-resonant firehose instability}\label{firehose}
The two CR streaming instabilities, resonant and non-resonant, despite
being rather different physical mechanisms, both rely on the non-zero,
first moment of the CR distribution, $\Delta_{1,m}$, i.e., the CR
current.  The anisotropic CR distributions, however, have non-vanishing
higher order harmonics, $\Delta_{l,m}$. The well-known example of the
instability due to the high-order anisotropy is the weakly magnetized
thermal background plasma where the gas pressure anisotropy can be
large.
The systems we consider are known to be unstable if the anisotropy is
large enough providing $\Ppar > \Pperp + B^2/4\pi$ (firehose
instability) or $\Pperp > \Ppar + B^2/4\pi$ (mirror instability)
\citep[e.g.,][and references therein]{tb97}, and can be important
in astrophysical shocks \citep[see, for example,][]{be87,bf07}. Since the
growth rate is $\propto k[(\Ppar - \Pperp - B^2/4\pi)/\rho]^{1/2}$,
the instability operates in hot, high $\beta$ plasmas.

In a strong supernova shock, the far
upstream plasma is cold unless the SNR
is located in a hot superbubble. Then, the free energy of the background
plasma pressure anisotropy is not an efficient driver of magnetic field
amplification. However, the CR pressure in the shock precursor close to the shock is
expected to reach a sizeable fraction of the shock ram pressure,
i.e., $P^{\rm CR} = \zeta \rho_a v_s^2/2$.
In this case, the anisotropy of
the CR pressure defined by $\Delta_{2,m}$ should play a role. In the
hydrodynamical regime, where the wavelength $2\pi/k$ is longer than the
CR mean free path, $\Lambda$, the growth rate of the firehose instability can
reach
\begin{equation}\label{fh}
\gamma^{fh}(k) \approx   \sqrt{\zeta\,|\Delta_{2,0}|/2}\,\, v_s\, k
\sim \sqrt{\zeta/2}\,\, \frac{v_s^2 k}{c}
\ .
\end{equation}
In this estimate, we assumed that
$\Delta_{2,0}\propto\Delta_{1,0}^2$. However, as we have pointed out
earlier, in some plasma flows with a complex structure of
quasi-regular magnetic fields, or in the flows with regular or
stochastic magnetic traps, one can expect much larger pressure
anisotropy with $\Delta_{2,0}\propto\Delta_{1,0}$ and, therefore,
higher growth rates.  The mirror instability would occur with a
growth rate similar to Eq.~(\ref{fh}) if the perpendicular component
of CR pressure dominates over the parallel pressure.

\subsection{CR-pressure gradient driven acoustic instability}\label{drury}
Drury and collaborators \citep[][]{dd85,df86} showed that acoustic
perturbations in a two-fluid plasma with a CR pressure gradient can grow
if the scale of the CR gradient is shorter than $D/a_0$, where $a_0$ is
the thermal gas sound speed and $D$ is the diffusion coefficient
\citep[see
also][]{chalov88,zankea90,md01,mds10}.
It is argued by \citet{md09} that this
acoustic instability can be important for DSA.
The growth rate of the
acoustic instability can be estimated as
\begin{equation}\label{acoust}
\gamma^{\rm ac} \approx \frac{|\nabla P^{CR}|}{2\rho_a\,a_0}
\end{equation}
for $k >\gamma^{\rm ac}/a_0$ \citep[see, e.g.,][for a
recent discussion]{mds10}.
Since $|\nabla P^{CR}| \sim \zeta v_s^3 /(\Lambda
c a_0)$, then $k_{\rm{min}} \sim \zeta (M_s^2\,v_s/c)/\Lambda$.
Thus,
the long wavelength fluctuations of $k \Lambda < 1$ that are the
most important to increase the maximum energies of CRs accelerated
by DSA, can grow only if $\zeta (M_s^2\,v_s/c) < 1$. This
condition is rather restrictive for young SNRs in the warm
interstellar medium. On the other hand, the short-scale fluctuations
amplified by the instability are important for heating the plasma upstream of the shock and can contribute to the ponderomotive force
discussed in section \ref{longwave}.

\citet{bjl09} proposed a DSA model in which stochastic magnetic fields
in the shock precursor are amplified through small-scale dynamo
effects. The solenoidal velocity perturbations that are required for the
dynamo to work are produced in the model through the interactions of the
pressure gradient of the CR precursor and the
density perturbations in the
inflowing fluid.

\subsection{Instabilities in partially-ionized plasmas}
The H$_\alpha$ emission from SN1006 and some other SNRs implies the
presence of a sizable fraction of partially-ionized plasma in the
circumstellar medium.
Using a kinetic description of CRs,
combined with a fluid description of the background plasma,
\citet{bt05}
and \citet{revilleea07} investigated the growth of hydro-magnetic waves
driven by CR streaming in the partially-ionized precursor environment of
a supernova remnant shock.  In this case, modifications of Ohm's law
in the
magnetized plasma can be important for the instability, as can
ion-neutral damping.
If the cosmic ray acceleration is efficient, large neutral fractions
are required to compensate the growth of the non-resonant mode. For
modest acceleration rates, the ion-neutral damping can dominate over
the instability of weakly driven modes even at modest ionization
fractions. In the case of a supernova shock interacting with a
molecular cloud, such as in IC~443, W28, W44, or W51C, the low
ionization in the ambient medium could limit the maximum energy of
accelerated particles since this is largely determined by the
upstream magnetic turbulence \citep[e.g.,][]{bceu00}. Observationally, SNRs interacting with molecular clouds are indeed found to be sources
of high-energy gamma-ray emission \citep[see
e.g.,][]{magic_ic443_07,hess_w28,fermi_ic443_10,cs10}. This emission is likely due to some combination of  pion-decay and relativistic bremsstrahlung
\citep[e.g.,][]{bceu00,uchiyamaea10}.

Charge exchange reactions in the downstream region, from
neutrals crossing the shock and interacting, could result in a fast, cold ion beam.
\citet{ott09} presented a linear analysis of collisionless plasma
instabilities between the cold beam and the hot downstream plasma.
They found that, under  SNR conditions, either the resonant
instability or the Weibel-type instability are growing. The authors
concluded that the mechanism may amplify the downstream magnetic
field to more than 100 $\mu$G, changing the shock structure and the
synchrotron spectra and profiles.
However, as we have pointed
out \emph{the turbulence amplification in the shock precursor} is the
main factor determining the maximum CR energy and
the efficiency of energy conversion to accelerated particles.


\section{Magnetic field amplification in diffusive shock acceleration}
The plasma instabilities in particle acceleration sources  discussed
above may result in magnetic turbulence and strong magnetic field amplification. The high efficiency of the acceleration means that nonlinear models are needed to predict the statistical
characteristics of the amplified magnetic fields and their
observational appearances. Diffusive shock acceleration models in supernova remnants imply
wide dynamic ranges for the fluctuations and the particle momentum spectra extending for more than four decades.
These wide dynamic ranges present a severe problem for the
direct use of particle-in cell plasma simulations and some
simplifications are required.

The nonlinear modeling of the short-scale Bell instability was done
using MHD simulations \citep[e.g.,][]{bell04,zp08,zpv08} assuming a
fixed CR current as an external parameter. The MHD approach is
justified since the CR current has only a weak response to
the short-scale fluctuations. These models studied the spectral
evolution in the short-scale range as well as the transformation of
the turbulence through the subshock.  The evolution of the Bell
modes downstream from the shock was addressed by \citet{plm06},
\citet{marcowithea06}, and \citet{mc10}.

\begin{figure*}
\includegraphics[width=0.8\textwidth]{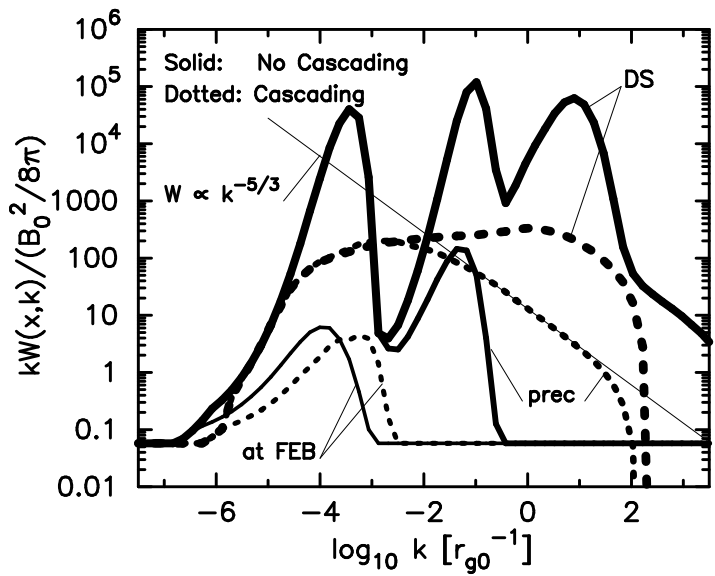}
\caption{Turbulence spectra, $k W$, at different locations relative
to the subshock position as simulated by Vladimirov et al. (2009).
The far upstream seed turbulence (at all $k$) is at the level
indicated by the horizontal lines.}\label{MC_spectra}
\end{figure*}

A nonlinear Monte Carlo  model of strong forward shocks in young
supernova remnants with efficient particle acceleration, where a
nonresonant instability driven by the CR current amplifies
magnetic turbulence in the shock precursor, was developed by
\citet{vbe09}.
Particle injection, magnetic field amplification, and the
nonlinear feedback of particles and fields on the bulk flow were derived
consistently in the Monte Carlo model as presented by
\citet{veb06,vbe08}. It was found that the shock structure depends
critically on the efficiency of turbulence cascading.  If cascading is
suppressed, MFA is strong, the shock precursor is stratified, and the
turbulence spectrum contains several discrete peaks. In
Fig.~\ref{MC_spectra}, simulated turbulence spectra, $k W$, are shown at
different locations relative to the subshock position.

In the solid curve, cascading is fully suppressed. In the dotted curve, the cascading from large to small scales is efficient and has the
differential form corresponding to Kolmogorov's model.
In the case with no cascading, the dissipation was assumed to be zero. With cascading, viscous dissipation was assumed.
 The peaks, apparent in the model with suppressed cascading, should
influence \syn\ X-ray images of SNR shells, allowing observational tests
of cascading and other assumptions intrinsic to the nonlinear model of
\nonres\ wave growth. In the next section, we discuss some models
of X-ray synchrotron imaging of SNR shells.

Two opposite limits of cascading were considered in the model by
\citet{vbe09} because our knowledge of turbulent cascading
in collisionless plasmas is still very limited. It is limited, in part, because of the
rather narrow dynamical range of the available nonlinear
simulations.
Information from direct measurements in the interplanetary medium can
help constrain models of turbulent cascading, but this is also
limited. The cascading of plasma wave spectra upstream of
interplanetary shocks driven by coronal mass ejection events were
analyzed by \citet{bamertea08}. They studied the competition between the
upstream wave generation by suprathermal protons accelerated at the
shock, and the cascading of wave energy in the inertial range of solar
wind turbulence. \citet{bamertea08} concluded that amplified solar wind turbulence
upstream of interplanetary traveling shocks is better described by
a Iroshnikov-Kraichnan-type cascade rather than the Kolmogorov-type.

The turbulent cascade is expected to be anisotropic, as is
the case in the interplanetary medium \citep[see e.g.,][and the
references therein]{matthaeusea96,hfo08,podesta09}. Turbulent
cascade dissipation will heat ions and electrons both in the upstream
and downstream regions of the shock. Since the magnetic field
amplification region in the shock precursor is of a finite size
$\sim (c/v_s) \Lambda$, the upstream plasma temperature depends strongly on
the heating rate and the plasma advection time. The cascade rate
$\tau_c^{-1}$ depends on the field amplitude, growing mode
polarization, and the plasma parameters.
If $\tau_c^{-1} (c/v_s^2) \Lambda <1$, the upstream plasma heating is
inefficient.
However, in the opposite limit with  strong
upstream turbulence, the plasma ion heating via an appropriate Landau
resonance can be strong if the plasma parameter
$\beta = a_0^2/v_a^2 > 1$.
Electron heating is dominant in the highly magnetized
upstream plasma if $\beta < 1$ \citep[e.g.,][]{qg99,howes08}. The
Monte-Carlo simulations of the nonlinear structure of CR modified
shocks by \citet{vbe08} demonstrated that the CR acceleration
efficiency and the shock modification are sensitive to ion heating by
the upstream turbulence. X-ray line emission from shocked shells of
SNRs can be used to constrain the DSA models
\citep[e.g.,][]{ellisonea10}.

\begin{figure*}
\includegraphics[width=0.6\textwidth]{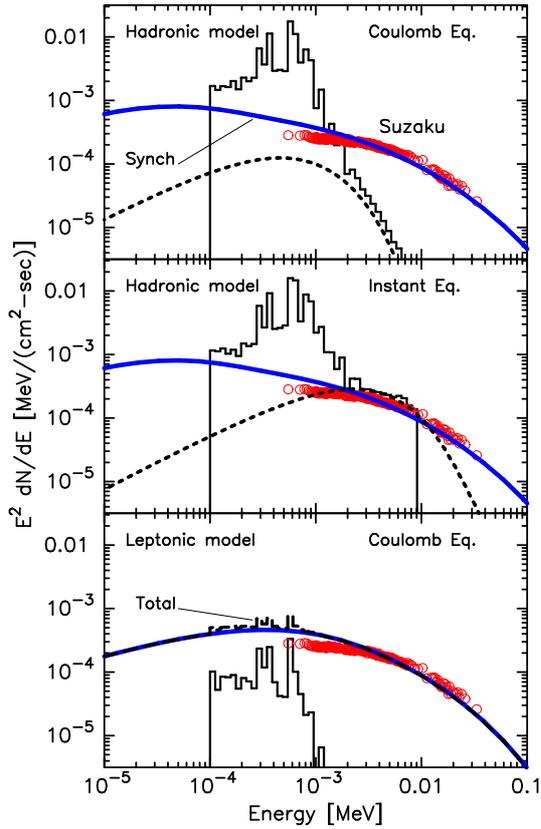}
\caption{The top two panels show fits to the {\it Suzaku} SNR J1713
observations
  with a hadronic model for both Coulomb and instant temperature
  equilibration but ignoring the X-line emission.  The blue (heavy wt.)
solid curve
  is the \synch\ continuum, the black solid curve is the thermal
  emission (only lines above $10^{-4}$\,MeV are included), and the
  dotted curve is the underlying bremsstrahlung continuum. The observed
  emission would be the sum (not shown in the top two panels) of the
  solid black and blue curves.  The bottom panel shows the leptonic
  model (with Coulomb equilibration) where parameters have been chosen to
  be consistent with the {\it Suzaku} observations.
For the hadronic model, the radiation intensity is multiplied by
0.95 to match the observations. For the leptonic model, a
normalization factor of 0.2 is required to match the observations.
The {\it Suzaku} data have been adjusted for interstellar extinction
so no extinction is applied to the model in this plot. See
\citet{ellisonea10} for full details.}\label{J1713_fig}
\end{figure*}

There is an emerging class of SNRs, such as SN1006,
 RXJ1713.72-3946, Vela Jr, and others, that are \emph{dominated} by
 non-thermal emission in X-rays, probably of synchrotron origin.
Such emission results from electrons and/or positrons that are accelerated to well above TeV
 energies. Models of supernova shells show that these leptons radiate in strong magnetic fields and the radiation losses produce a marked cut off in the lepton spectrum.
Figure~\ref{J1713_fig} \citep[from][]{ellisonea10}
 shows the X-ray emission from RXJ1713.72-3946 as observed by the Suzaku
 spacecraft. This figure, and the models given in \citet{ellisonea10},
 indicate that the lack of thermal
X-ray emission lines can be important for
 determining the origin of the non-thermal emission and the strength of the magnetic fields in the SNR.
Since the amplified magnetic
fields in DSA must be fluctuating, we next discuss the effect of
fluctuations on the synchrotron images of SNRs.

\section{Synchrotron images, spectra and polarization in supernova
shells}\label{synchr1}
Non-thermal emission in many sources of
interest such as SNRs, gamma-ray bursts, and AGNs
originates from synchrotron radiation of ultra-relativistic
particles in turbulent magnetic fields. The effect of a random
magnetic field on synchrotron images, emission spectra, and
polarization can be very substantial.

\subsection{Synchrotron mapping of twinkling supernova shells}\label{maps}
It is instructive to illustrate first the effect of intermittency
caused by the random magnetic field in terms of a power-law electron
spectrum with spectral index $\Gamma$. The synchrotron emissivity
$\tilde{I}({\bf r},t, {\nu}) \propto B^{(\Gamma + 1)/2}$ shows that
the local emissivity is relatively very high for large $B$ and large
$\Gamma$ (a large $\Gamma$ implies a steep spectrum).
Therefore, the high-order statistical moments of the field
dominate the synchrotron emissivity in the spectral cut-off regime and it is possible for a single strong local field maximum to produce a feature
(dot or clump) that stands out on the map even after integrating the
local emissivity over the line of sight. In lower energy maps, the
contribution of a single maximum can be smoothed or washed out by
contributions from a number of weaker field maxima integrated over
the line of sight.
A high-energy map, however, can be highly intermittent
because the synchrotron emissivity depends on high-order moments of
the random magnetic field at the cut-off frequency regime. The
electron/positron spectra in the cut-off regime are usually
exponential, and the exact shapes of the spectra dominated by
synchrotron/Compton losses depend on the diffusion coefficient.
A power-law approximation can only apply for a narrow electron energy
range in the cut-off regime, where the effective spectral index $\Gamma$
is increasing with the electron energy.

In order to calculate synchrotron emission maps
and spectra,  \citet{bue08} modeled a
system of finite size, filled with a random magnetic field and electron spectra simulated with a DSA model accounting for
synchrotron/Compton losses \citep[see][]{bceu00}.
A forward shock of spherical geometry was
assumed in the image construction.
%
The random field was composed of a
superposition of magnetic fluctuations (plane waves) with random
phases and a given spectrum of amplitudes.
\emph{Non-steady},
localized structures (dots, clumps and filaments), in which the
magnetic field reaches exceptionally high values, typically arise in
the statistically stationary random field sample. These non-steady
magnetic field concentrations dominate the synchrotron emission
(integrated along the line of sight) from the highest energy
electrons in the cut-off regime of the electron
distribution, resulting in an
intermittent, twinkling, clumpy appearance even for a steady
distribution of ultra-relativistic electrons. The spectral and
temporal evolution of these high intensity synchrotron events, and the
synchrotron images they produce,
vary drastically at different photon energies.

\begin{figure}[t]
\includegraphics[width=1\textwidth]{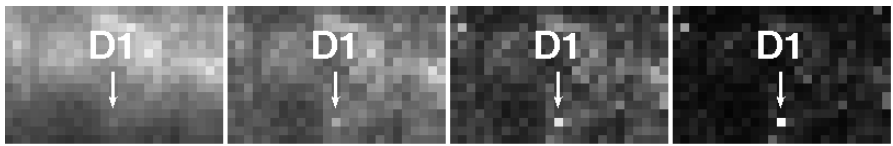}
\caption{Emission in the vicinity of a bright synchrotron dot D1 at
different photon energies (see Bykov et al. 2008 for details). The
four panels show, from left to right, $\nu^2\cdot I({\bf
R_{\perp}},t, {\nu})$ maps of synchrotron emission at 0.5, 5, 20 and
50 keV, respectively.}\label{clump}
\end{figure}

In Figure~\ref{clump} the synchrotron emission of a small region
around a dot labeled D1 is shown for emission at 0.5, 5, 20 and
50 keV. There are observable differences in the synchrotron maps
indicating that some features are bright at high energies and much
less prominent at lower energies and vice versa.
The physical reason for this difference comes from the fact that in
the high-energy cut-off region of the electron spectrum, the local
synchrotron emissivity depends strongly on the local field value as
$B^{(\Gamma + 1)/2}$, where the effective index $\Gamma$ is
increasing with the photon energy in the cut-off spectral regime.

The simulated light curves of X-ray clumps may explain those
observed in X-ray images of some supernova remnants by
\citet{uchiyamaea07} and \citet{pf08}. It is important to note that if the clumps are due to
magnetic field intermittency, the time scale of their variation is not reflecting the ultra-relativistic electron loss time.
The distinct characteristic of the modeled synchrotron emission is
that its strong intermittency results directly
from the exceptionally
high magnetic field amplifications that randomly occur.
%
The peaks in synchrotron emission maps occur due to
high-order moments of the magnetic field probability distribution
function (PDF). Even if the PDF of projections of the local magnetic
field are nearly Gaussian, the corresponding PDF of synchrotron
peaks, simulated for a spatially homogeneous relativistic particle
distribution, has strong departures from the Gaussian at large
intensity amplitudes.
This is because of the nonlinear dependence of the synchrotron
emissivity on the local magnetic field in the high-energy cut-off
regime of the electron spectrum.  Intermittency of this nature, with very different
appearances, is rather a
common phenomenon in stochastic media \citep[see e.g.,][]{zeldovichea87}.

The models by \citet{bue08} and \citet{bubhk09} make a few basic
predictions. One is that the overall efficiency of synchrotron
radiation from the cut-off regime in the electron spectrum can be
strongly enhanced in a turbulent field with some value of
$\sqrt{\langle B^2\rangle}$, compared to emission from a uniform
field, $B_0$, where $B_0 = \sqrt{\langle B^2\rangle}$.
The second is that strong variations in the brightness of small
structures can occur on time scales much shorter than variations in
the underlying particle distribution. The variability time scale is
shorter for higher energy synchrotron images.
An estimate of the time scales of these intensity variations is
consistent with the rapid time variability seen in some young SNRs
by \emph{Chandra}. The strong energy dependence predicted may be
important for the future missions \emph{NuSTAR}, \emph{Astro-H} and
others that will image SNR shells up to 50 keV.

\subsection{Synchrotron X-ray Polarization in DSA}
Synchrotron radiation in a regular, quasi-homogeneous magnetic field is polarized \citep[e.g.,][]{gs65}. It has long been known that random
directions of magnetic fields, in addition to Faraday rotation, may
strongly reduce the \emph{average} polarization of synchrotron
emission sources \citep[e.g.,][]{westfold59, cs86,sp09}. This
explains the relatively low polarization frequently observed for
radio synchrotron sources. However, as it was shown by
\citet{bubhk09}, the turbulent magnetic fields that reduce the
\emph{average} polarization can  result in highly polarized,
intermittent, patchy structures potentially observable in high
resolution X-ray images.
%
%
In terms of
a power-law electron spectrum with spectral index $\Gamma$, the
degree of polarization is given by $\tilde{\Pi} \approx (\Gamma +
1)/(\Gamma + 7/3)$ and, therefore,  $\tilde{\Pi}$ increases with
$\Gamma$. Thus the synchrotron images will be highly polarized at
high energies.

In order to construct maps of polarized synchrotron emission from
SNR shells, it is convenient to use the local densities of the
Stokes parameters \citep[see e.g.,][]{gs65}. Because of the additive
property of the Stokes parameters $\tilde{I}, \tilde{Q},\tilde{U},
\tilde{V}$ for incoherent photons, we can integrate these weighted
with the distribution function of radiating particles over the line
of sight across the source. The degree of polarization is then
determined in a standard way as $\Pi = \sqrt{Q^2 + U^2 + V^2}/I$.
Starting from the simulated random magnetic-field, \citet{bubhk09}
have constructed maps of {\sl polarized} X-ray emission of SNR
shells. These are highly clumpy with polarizations up to 50\% for
energetic
$>$TeV electrons in the cut-off regime.

\begin{figure}
\centering {
 \rotatebox{0}{
{\includegraphics[height=5.6cm, width=7.5cm]{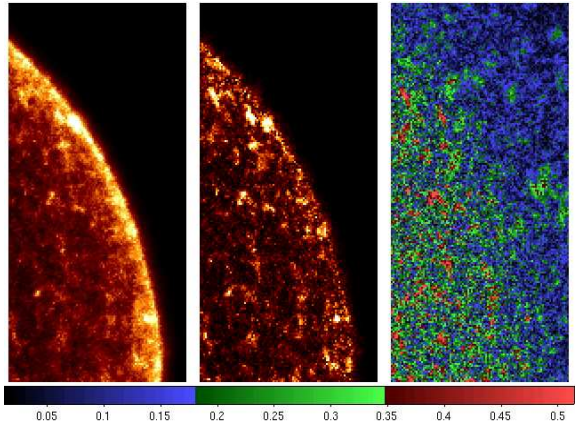} }}}\caption
{Simulated maps of polarized synchrotron
  emission in a random magnetic field at 5 keV.
  Intensity, $\nu^2\cdot I({\bf
  R_{\perp}},t, {\nu})$, is shown with a linear color scale in the left panel.
  The central panel  shows the product of intensity and polarization
  degree. The right panel shows the degree of polarization
  indicated by the color bar.
The stochastic magnetic field sample has $\sqrt{\langle B^2\rangle}
= 3\times 10^{-5}$ G and spectral index of magnetic fluctuations is
$\delta = 1.0$  (from Bykov \etal (2009)).
  }\label{polar}
\end{figure}


The detection of polarization in X-ray sources as faint as 1 milliCrab
is an aim of the \emph{Gravity and Extreme Magnetism Small Explorer
(GEMS)} mission, recently selected for flight in 2014 by NASA. It will
make a sensitive search for X-ray polarization using thin foil mirrors
and Time Projection Chamber polarimeters with high efficiency in the
2-10 keV band \citep{gems10}. The polarimeters under consideration, like
\emph{XPOL} aboard the planned \emph{IXO}
mission\footnote{http://ixo.gsfc.nasa.gov/science/performanceRequirements.html} and
 \emph{POLARIX} proposed to measure X-ray polarization with an
angular resolution of about 20 arcsec in a field of view of $15 \times
15$ arcmin with the minimum detectable polarization of 12\% for a
source of  1 mCrab within 100 ks of observing time
\citep{costaea10}, have good perspectives to detect polarized
synchrotron emission from SNR shells.

The intermittent appearance of the polarized X-ray emission maps
of young SNR shells can be studied in detail with
imagers of a few arcsecond  resolution, though even arcmin
resolution images can provide important information as is
illustrated in Fig.~\ref{polar} \citep[see][]{bubhk09}.
The polarized emission clumps of arcsecond  scales
are time variable on a year or longer (depending on the observed
photon energy, magnetic field amplification factor, and the plasma
density in the shell) allowing for rather  long exposures even in
the hard X-ray energy band. Hard X-ray observations in the spectral
cut-off regime are the most informative to study the magnetic
fluctuation spectra and the acceleration mechanisms of
ultra-relativistic particles.

\section{Magnetic field amplification in star forming galaxies}
The origin and evolution of magnetic field structures in galaxies is
an important unsolved problem  \citep[e.g.,][]{rss88,rees06,kz08}.
Radio observations of the diffuse polarized radio emission from the
disks of some spiral galaxies, as reviewed by \citet{beck08},
revealed  large-scale coherent magnetic patterns that could be
generated by dynamos, and dynamo models have been developed to
explain the main observational features of the global magnetic
fields of spiral galaxies \citep[e.g.,][]{rss88,beckea96,bs05}. In
most galaxies, however, the field structure is more complicated.
For example, radio synchrotron observations
of nearby galaxies reveal dynamically important magnetic fields of
10-30 $\mu$G total strength in the spiral arms. These spiral arm fields, however, tend to have random
orientations, while ordered fields
(observed in radio polarization) are strongest in interarm regions
and follow the orientation of the adjacent gas spiral arms.

A variety of observations \citep[see][for a review]{widrow02}  suggest
that magnetic fields are present in all galaxies including ellipticals and in galaxy clusters. Furthermore, it remains to be understood, in
the context of the dynamo models, why well organized fields of
surprisingly high strengths are observed in normal galaxies when
the Universe was much younger than its present age \citep[see
e.g.,][]{bernetea08,wolfeea08}.

\subsection{Magnetic fields in irregular galaxies}
Another issue to be addressed concerns the magnetic fields of dwarf
irregular galaxies. Studies of magnetic fields in nearby dwarfs, as
discussed recently by \citet{chyzy10}, revealed typically weak fields
with the mean value of the total field strength about three times
smaller than in normal spirals. The slow rotation rate in many
irregular galaxies makes standard dynamo-type  amplification
of the magnetic field unlikely.
On the other hand, dwarfs with
stronger fields are associated with violent star-forming activity
and these tend to be  more massive and evolved systems. Their magnetic fields
are thought to be regulated mainly by the surface density of the
galactic star-formation rate \citep[e.g.,][]{chyzy10}.

Star forming
activity with clustered supernova explosions in superbubbles
observed in galaxies \citep[see e.g.,][]{heiles90} may affect the
mean magnetic field as suggested by \citet{Ferriere92}.
Moreover, numerical magnetohydrodynamical models by
\citet{Siejkowskiea10} of cosmic-ray driven dynamos in the
interstellar medium of an irregular galaxy indicated the possibility
of magnetic field amplification.
The process of cosmic-ray acceleration may play a role in generating
galactic magnetic fields at different stages of cosmic evolution.
Non-resonant mechanisms of magnetic field amplification allow  field
enhancements on scales much larger than the gyroradii of the
ultra-relativistic particles accelerated in the sources. Such fields
amplified by the CR current and CR pressure anisotropy will have
scales well above a parsec in the star forming regions, and well
above a kiloparsec in clusters of galaxies. Observations of
supernova remnants and galactic CRs suggest that there is a high
efficiency of conversion of the kinetic power of supernova shocks
into CRs. Therefore, the specific mechanisms discussed above may
convert a sizable fraction of the power released by SNRs into
intermediate scale magnetic fields that are much larger than stellar
scale fields. Some large-scale dynamo mechanism may also be
necessary to produce the largest scale coherent fields seen in
spiral galaxies.

\subsection{Magnetic fields in the first generation Pop III supernova remnants}
Simulations of
nonlinear cosmic structure formation by
\citet{abelea02} and \citet{xuea08} showed that the high-redshift analog of a
molecular cloud is able to produce individual massive primordial
stars, offering a natural explanation for the absence of purely
metal free low mass stars in the Milky Way.
The production of  small-scale dynamos by
turbulence created by accretion shocks during gravitational collapse
may amplify initially weak magnetic fields in the protostellar cloud
in the formation process of the first stars and galaxies
\citep[i.e.,][]{schleicherea10}. Massive stars exploding as supernovae
provide energy, momentum, entropy, and metals at the very earliest  stages of
galactic evolution. Observations of high-redshift gamma-ray bursts
likely indicate that relativistic particles and magnetic fields
are produced during the collapse of massive primordial stars.

An important problem is whether there is a way to transfer a
significant fraction of the high kinetic energy of the supernova
ejecta into the magnetic fields of the remnant. The questions to be
addressed are the following. \begin{itemize}
\item (i) Can
{\it collisionless} shocks be created by the supersonic ejecta in the
very weakly magnetized plasma of $\beta
>> 1$?
\item (ii) Can relativistic particles be accelerated in Pop III
supernova remnants?
\item  (iii) Are there effective mechanisms of magnetic field
amplification in Pop III supernova remnants?
\end{itemize}

Two-dimensional particle-in-cell simulations performed by
\citet{kt10} to study weakly magnetized perpendicular shocks
demonstrated the development of a {\it collisionless} shock with  $\beta \approx 26$ (and an \alf Mach number of about 100).
They showed that current filaments form in the foot region of the
shock due to the ion-beam-Weibel instability (or the ion
filamentation instability) and these filaments generated a strong magnetic
field there. Strong fluctuating field generation is necessary in
the standard paradigm of a collisionless shock since the only way
to decelerate and thermalize the cold upstream flow is through the
electromagnetic fields.
In the downstream region, these current filaments also generated a
tangled magnetic field that was about 15 times stronger than the
upstream magnetic field. The shock simulations by \citet{kt10}
reproduced well the Rankine-Hugoniot relations and indicated a
substantial damping of the strong Weibel-type field fluctuations in
the shock ramp (in the case of $\beta \approx 26$, the magnetic
field contribution in the Rankine-Hugoniot relations is small). The
authors  noticed that a fraction of the ions are accelerated
slightly on reflection at the shock.

If the temporal and spatial scales of the simulation were increased,
it is expected that this  fraction of reflected superthermal
particles would continue to  be accelerated by the Fermi-type DSA
process. In this case, the magnetic field would continue to be
amplified in a symbiotic relationship where the magnetic turbulence
required to scatter and accelerate the CRs is created by the
accelerated CRs themselves  \citep[e.g.,][]{bell04,vbe08,boe10}.
If this amplification operates as expected, a sizable fraction of
the energy released by the supernova explosion can be transferred to
energetic particles even with the initially weak magnetic field
fluctuations of $B_0 \sim$ 0.1-1 nG expected in the protostellar
cloud.


The largest scale magnetic fluctuations amplified by the long
wavelength non-resonant instabilities (that are discussed above in
\S\ref{mfa}) can approach the scale of the shock. Therefore,
the
supernova remnant and its vicinity, with a scale size of about 30 pc,
can be filled with a magnetic field of strength  $\sim$ 0.1
$\mu$G generated by instabilities driven by CR acceleration.
The longest scale fields generated by the non-resonant CR
instabilities in the shock
\emph{precursor}
are in the collisional plasma regime since their wavelengths are
longer than the Coulomb mean free path of the ambient thermal
plasma. The long-wavelength fields should have, by many orders of
magnitude, longer damping times than the short-lived
Weibel-type fluctuations that are located just in a very narrow
vicinity of the shock jump.
The magnetic fields amplified in the SNR
shock precursor
can be considered as mesoscale fields since their scales are much
larger than the stellar field sizes and larger than the CR
gyroradii, but are  below the observed largest scale coherent
galactic magnetic fields in spiral galaxies presumably generated by
galactic dynamos or by a global galactic CR current.
The mesoscale fields can
be the dominant fields in irregular galaxies and can be spread by the
galactic winds through galactic groups and clusters. Furthermore,
CR-driven winds in high-redshift galaxies may have important implications for the
metal and magnetic field enrichment of the
 intergalactic medium \citep[see e.g.,][]{samuiea10,dt10}.
In contrast, magnetic fields much lower than expected from CR instabilities
were predicted in MHD simulations of the Biermann battery mechanism
in the first generation of
supernova remnants \citep[i.e.,][]{hanayamaea09}.


Magnetic field amplification by CRs produced by the first supernova explosions during
the re-ionization epoch of the universe was discussed by
\citep{mb10}. The CRs escape into the intergalactic medium carrying
an electric current that results in magnetic field amplification.
The authors found that magnetic fields are robustly generated
throughout intergalactic space at a rate of 10$^{-17}-10^{-16}$
Gauss/Gyr, until the temperature of the intergalactic medium is
raised by cosmic
re-ionization. In another shock-related process, \citet{ryuea09} discussed a scenario
in which turbulent-flow motions are induced via the cascade of vorticity generated at cosmological shocks during the formation of
large-scale structure. It was argued that this turbulent dynamo model might provide average magnetic field strengths of a few
microgauss inside clusters and groups, approximately 0.1 $\mu$G
around clusters and groups, and approximately 10 nG in filaments.

%
\section{Discussion}
Magnetic fields, both regular and stochastic, play a key
role in charged particle accelerators providing  both particle
confinement and energy gain.
The  enormous energy
release of gravitational (AGN, GRB, SNRs, clusters of galaxies, etc.)
or thermonuclear origin (Type Ia SNRs, etc.), produces
collisionless shocks and accelerates particles in many sources.
It has long been
known that strong shocks can only form in collisionless plasmas if
magnetic turbulence is generated and some thermal particles are
injected and accelerated to superthermal energies.
The strong interrelation between shock formation and structure,
magnetic turbulence, and energetic particles makes understanding the
self-consistent production of magnetic turbulence in \DSA\ both
essential and difficult.

While a great deal of progress has been made in this field, important
challenges remain. In this brief review, we have discussed some
recent work including resonant streaming and non-resonant CR-current
driven instabilities. An important example of this is the Bell mechanism
that efficiently amplifies short-scale turbulence by the CR current.
Magnetic turbulence amplification may govern the
nonlinear structure of CR-dominated flows, heat the upstream
plasma, enhance synchrotron radiation, and limit high-energy electron
energies. Furthermore, this mechanism might operate hand-in-hand with an
instability for producing long-scale turbulence. The long-scale
turbulence can be responsible for the highest energy particles
accelerated by DSA. The basic DSA prediction that the highest energy accelerated ions can have a hard spectrum implies that the highest energy CRs may
contain a sizeable fraction of the shock ram pressure. This implies
that a consistent model of magnetic fields in the
highly supersonic flow with accelerated particles is nonlinear and
multi-scale.

The field of \NL\ shock acceleration  is evolving rapidly with several prominent areas of
active study. The long-wavelength instabilities discussed in
Section~\ref{mfa} are yet to be combined with the resonant and
short-wavelength instabilities in a nonlinear shock simulation. Such
a combination will determine the relative importance of each and how
the overall shock structure is influenced by them.
The instabilities produced by escaping cosmic rays are only just now
starting to be explored. Only with a consistent model combining the
turbulence produced by escaping cosmic rays on the escaping cosmic
rays themselves, can a reliable estimate of the maximum particle
energy a given shock can produce be obtained.

In principle, the plasma physics of shock structure, magnetic
turbulence, particle injection and acceleration, indeed, all aspects
collisionless plasmas, can be determined with particle-in-cell (PIC)
simulations. A great deal has been done in this field and we have
mentioned a small fraction of this work in Section~\ref{bell}. The
computational difficulties in producing PIC simulations with
parameters close enough to systems such as SNRs are severe and we
see future progress coming from synthesis of PIC models with
semi-analytic and Monte Carlo techniques.

We have also discussed some recent work on the observational
consequences of magnetic turbulence and \syn\ emission. The important result is that  fluctuating magnetic fields strongly modify the
spectra of synchrotron emission in the case of steep electron spectra
that are typical for X-ray emitting, TeV particles in young
SNRs. X-ray images of synchrotron dominated SNR shells are predicted
to be highly intermittent with polarized twinkling structures of
different scales if the strong amplification of magnetic turbulence
is indeed operating in DSA.

The current generation of X-ray telescopes -- \emph{Chandra} with its
  superb angular resolution and \emph{XMM-Newton} with its high
  sensitivity -- has uncovered a new world of synchrotron structures in
  young SNR shells.  Further progress is certain to come from improved
  polarization and timing observations of SNRs at X-ray energies where
  the Faraday rotation is not essential and characteristic time scales
  are short enough. Observations at GeV-TeV energies are becoming more
  sensitive and theories of the highest energy cosmic rays must be
  improved accordingly. In all aspects of theory and observations, the
  turbulent magnetic field is the complicated glue tying things
  together.

\begin{acknowledgements}
We thank the referee for a constructive comment.  A.M.B. was
supported in part by RBRF grants 09-02-12080  and by the
RAS Presidium Programm, and the Russian government grant
11.G34.31.0001 to Sankt-Petersburg State Politechnical University.
He performed some of the simulations at the Joint Supercomputing
Centre (JSCC RAS) and the Supercomputing Centre at Ioffe Institute,
St. Petersburg. D.C.E acknowledges support from NASA grants
ATP02-0042-0006, NNH04Zss001N-LTSA, and 06-ATP06-21.

\end{acknowledgements}

\bibliographystyle{svjour}
\bibliography{mf_1}

\end{document}